# A Comprehensive Study of Transfer Learning, Resolution Reduction, and Multi-View Classification for Breast Cancer Detection in Mammograms

Daniel G. P. Petrini | Hae Yong Kim

Department of Electronic Systems Engineering, Polytechnic School, University of São Paulo, Sao Paulo, Brazil

**Correspondence:** Daniel G. P. Petrini (dpetrini@alumni.usp.br)

## ABSTRACT

Mammography, an X-ray–based imaging technique, remains central to the early detection of breast cancer. Recent advances in artificial intelligence have enabled increasingly sophisticated computer-aided diagnostic methods, evolving from patch-based classifiers to whole-image approaches and then to multi-view architectures that jointly analyze complementary projections. Despite this progress, several critical questions remain unanswered. Although individual aspects of CNN-based mammogram classification—such as patch pretraining, backbone transferability, and multi-view fusion—have been studied separately in prior work, no unified comparison has examined them under controlled conditions across datasets of contrasting image quality. In this study, we close that gap through a systematic empirical investigation organized around five interrelated questions: (1) the role of patch classifiers, (2) the transferability of natural-image-trained backbones, (3) the advantages of "learn-to-resize" over conventional downscaling, (4) the contribution of multi-view integration, and (5) the robustness of findings across varying image quality. Questions (3) and (5) are addressed here for the first time, while questions (1), (2), and (4) are re-examined within a unified protocol that reveals dataset-quality dependencies invisible to prior single-dataset studies. Beyond benchmarking, our experiments demonstrate clear performance gains over prior work. For the CBIS-DDSM dataset, we improved single-view AUC from 0.8153 to 0.8343, and multiple-view AUC from 0.8483 to 0.8658. We also introduce a comparative protocol that places single- and multi-view classifiers on the same evaluation footing—by aggregating per-view single-view predictions into per-exam scores—enabling a statistically valid comparison between the two; using this protocol, we measured a 0.0217 AUC gain from extending single- to multi-view analysis on CBIS-DDSM. On the complete VinDr-Mammo dataset, the multiple-view approach further improved results, achieving a 0.0492 AUC increase over single view and reaching 0.8511 AUC overall. These results establish new state-of-the-art benchmarks, providing clear evidence of the advantages of multi-view architectures for mammogram interpretation. Beyond performance, our analysis offers principled insights into model design and transfer learning strategies, contributing to the development of more accurate and reliable breast cancer screening tools. The inference code and trained models are publicly available at https://github.com/dpetrini/multiple-view.

## 1 | Introduction

Breast cancer remains the most prevalent cancer among women globally, accounting for approximately 23.8% of all new cancer cases diagnosed in 2022 [1]. Mammography, an imaging technique based on X-rays, is currently the gold standard for early breast cancer detection. However, interpreting mammograms requires considerable expertise, and despite experienced radiologists utilizing computer-aided detection and diagnosis (CAD) systems, errors remain prevalent.

For instance, Bahl and Do [2] reported a sensitivity of only 87% in physician-interpreted mammograms, with an interval cancer rate—cancers diagnosed following a negative mammographic result—of approximately 30%. These figures highlight significant limitations in traditional diagnostic methods and emphasize the potential benefit of integrating AI-based CAD systems to reduce errors.

Early applications of convolutional neural networks (CNNs) to mammogram classification, such as the work of Kooi et al. [3], highlighted the challenges of direct transfer from natural image tasks. The subtle and localized appearance of cancerous lesions, combined with the high-resolution, single-channel nature of mammograms, limits the effectiveness of conventional CNNs and underscores the need for specialized methodologies.

To address these challenges, Shen et al. [4] proposed a two-stage transfer learning approach. Initially, they trained a "patch classifier" using ImageNet-pretrained CNNs [5] to categorize small mammogram regions (patches) into lesion types. Subsequently, this trained patch classifier initialized a "single-view whole-image classifier," which produced per-image cancer probability. Petrini et al. [6] expanded this approach by introducing a third stage that simultaneously analyzed both medio-lateral oblique (MLO) and cranio-caudal (CC) mammographic views, significantly improving accuracy when tested on public datasets.

Despite these advancements, it remains uncertain whether CNN architectures that achieve top performance on natural image benchmarks (e.g., ImageNet) will consistently outperform other models when applied to mammograms.

While prior work has individually examined patch-based pretraining [4, 6], the transferability of ImageNet backbones [3, 7], and multi-view fusion [6, 8–10], these investigations have been conducted in isolation, with heterogeneous datasets, splits, and evaluation protocols, making it difficult to draw principled conclusions about when and why each design choice helps. In this paper, we contribute a unified experimental framework that reexamines these questions alongside two previously unaddressed ones: the suitability of the "learn-to-resize" technique for mammograms, and how all of the above design decisions interact with image quality. Our contributions are therefore: (i) the first evaluation of learn-to-resize for mammogram classification, showing it offers no benefit over conventional downscaling; (ii) a cross-dataset analysis that reveals how image quality fundamentally mediates the effectiveness of each design choice, for example, patch-based pretraining is beneficial for high-quality digital mammograms but not for digitized analog ones, a finding no single-dataset study could establish; (iii) controlled, reproducible validation of patch pretraining, backbone selection, and multi-view fusion under a single protocol, including a per-exam aggregation procedure that enables statistically valid comparison between single- and multi-view classifiers, a comparison that is methodologically ill-posed in much of the prior literature; and (iv) new state-of-the-art benchmarks, improving CBIS-DDSM single-view AUC from 0.8153 to 0.8343, CBIS-DDSM multi-view AUC from 0.8483 to 0.8658, and reaching 0.8511 multi-view AUC on the full VinDr-Mammo dataset.

The remainder of this paper is organized as follows. Section 2 reviews the public mammography datasets and prior work on single- and multi-view classification, and then presents our experimental methodology in detail. Section 3 reports and discusses the results obtained on the CBIS-DDSM and VinDr-Mammo datasets, addressing each of the five research questions in turn. Section 4 summarizes our main findings and concludes the paper.

## 2 | Materials and Methods

For breast cancer classification, machine learning studies may rely on proprietary mammography datasets, as in Wu et al. [11] and McKinney et al. [12], or on publicly accessible datasets. In this work, we exclusively utilize open-access mammographic repositories. In the next topic, we summarize the key characteristics of public datasets available, and justify our final selection. We then provide a literature review of techniques for breast cancer analysis, covering single-view and multi-view architectures. We conclude this section by presenting our experimental methodology in detail, including the research questions guiding the study, the dataset splits and training protocol, and the procedures used to build and compare patch-based, single-view, resolution-reduced, and multi-view classifiers across both datasets.

### 2.1 | Public Datasets

There are two principal types of mammography: screen-film mammography (SFM, conventional film-based) and full-field digital mammography (FFDM). To process SFM digitally, film images must first be scanned—further degrading their quality—whereas FFDM is inherently digital. A simple visual comparison of scanned SFM versus FFDM confirms that the latter offers substantially superior image quality, as in Figure 1. Accordingly, it is evident that training and evaluating an AI system on native digital mammograms will yield better results than using digitized analogue films.

To contextualize our dataset selection, we first survey the principal public mammography repositories available for deep learning research, summarizing their key characteristics and limitations, before justifying our final choice.

Fair comparison of different computer-aided detection (CAD) systems for breast cancer requires large public mammography datasets with reliable ground-truth labels. Ideally, such datasets should also include FFDM images, since they better reflect modern clinical acquisition. However, early publicly available mammography repositories consist of low-quality digitized SFM images: the Digital Database for Screening Mammography (DDSM) [13], the Mammographic Image Analysis Society database (MIAS) [14], and the Image Retrieval in Medical Applications project (IRMA) [15]. DDSM is the largest, comprising 2620 exams with normal, benign, and malignant cases verified by pathology. The InBreast dataset [16] contains true FFDM images but includes only 115 cases (410 images), which is too small for deep-learning applications.

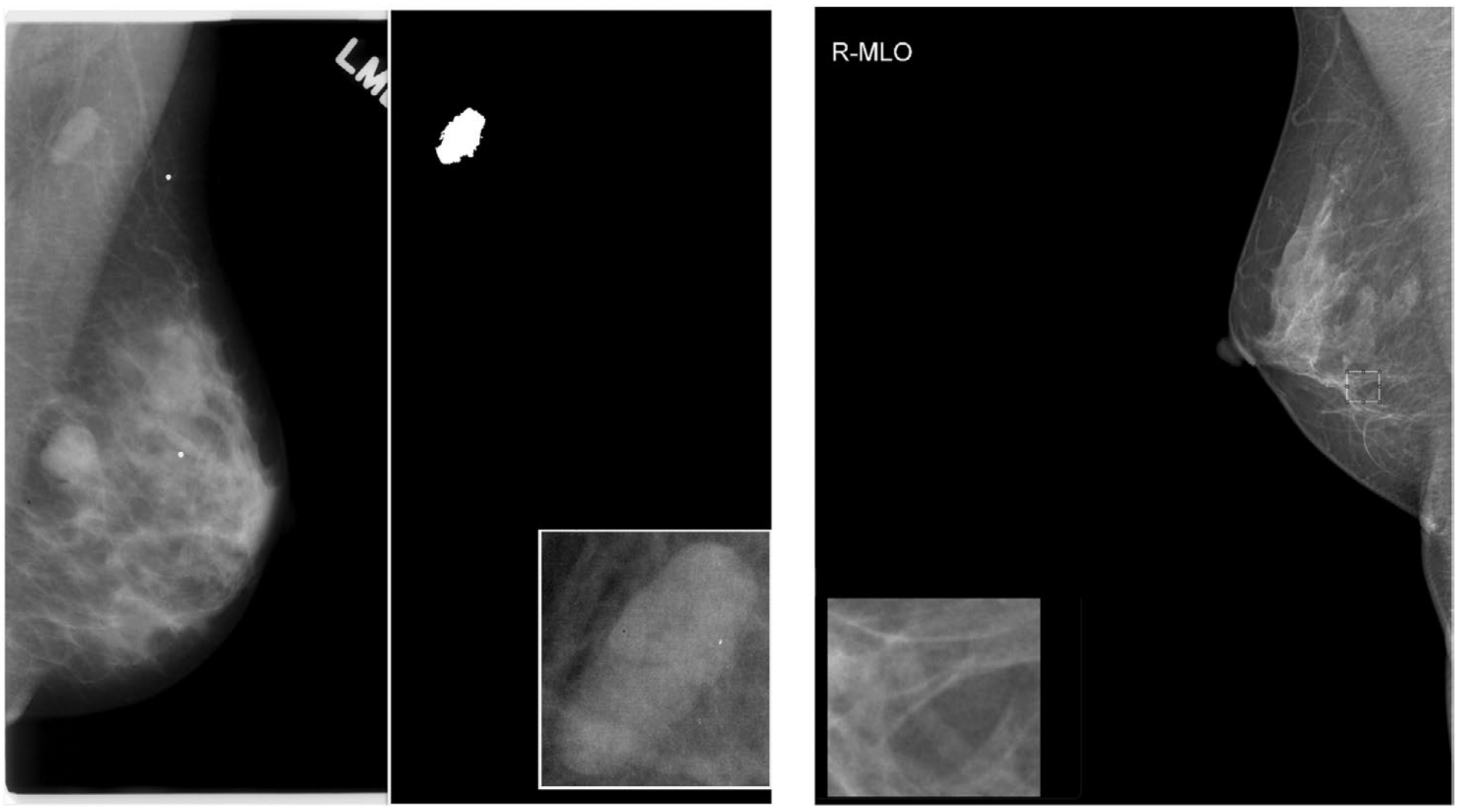

(a) (b)

**FIGURE 1** | Samples of public dataset used in this work. CBIS-DDSM film-based capture on the left and VinDr-Mammo digital capture on the right. (a) Mammogram in the left, and mask in the right. The mask indicates the location of the lesion. In rightmost bottom, the patch of mass lesion. All mammograms on dataset have these three types of images. (b) Mammogram with a box indicating the location of lesion. In leftmost bottom, the patch of lesion. Only full image present in dataset, patch was extracted from metadata with coordinates, available in dataset.

The CBIS-DDSM [17] is a curated and enhanced subset of the original DDSM, specifically tailored to support research applications. It comprises 3103 mammographic images, benign or malignant, in this case confirmed by biopsy, and includes detailed lesion annotations to facilitate automated analysis.

The CSAW-M dataset [18], developed at the Karolinska Breast Center in Sweden, comprises 10 020 mammograms at 632 × 512-pixel resolution, annotated by at least one radiologist with a "masking" difficulty score rather than benign/malignant labels, reflecting lesions obscured by surrounding tissue.

The KAU-BCMD [19] from King Abdulaziz University in Saudi Arabia includes 1416 exams (four views each, totaling 5662 images) labeled by BI-RADS rather than gold-standard biopsy; 205 of these exams also include ultrasound images (BI-RADS 0), and the public download currently provides 2378 mammograms at 1267 × 578 resolution (version 10).

VinDr-Mammo [20] is a full-field digital mammography (FFDM) collection of 5000 four-view exams (20 000 images) with BI-RADS assessments, breast density, and lesion annotations (masses, calcifications, asymmetries, etc.); each exam was independently double-read with arbitration for discordance, and only BI-RADS 3–5 exams include lesion marks.

EMory BrEast Imaging Dataset (EMBED) [21] is another large scale FFDM dataset, comprising 3.65 million screening and diagnostic 2D and DBT mammograms, from racially diverse backgrounds. A portion of this dataset is available for download. It contains biopsy and ROIs (regions of interest) pointing to the lesions.

The RSNA dataset [22] was the subject of a recent Kaggle data challenge and contains 54 713 digital mammograms from almost 8000 patients, but lacks marked regions of interest (ROI). It has indication of Bi-Rads, biopsy and has around 2.1% of positive cases.

The CMMD [23] Chinese Mammography Database contains 3712 DICOM mammograms (2294 × 1914 pixels, 8-bit) from 1775 patients, split into 2214 images with biopsy-confirmed benign or malignant labels and 1498 images with available molecular subtype information, but also lacks marked regions of interest (ROI).

NL-Breast-Screening [24], is a recent Canadian population-screening FFDM dataset of 5997 four-view exams, biopsy-confirmed, with positive, false-positive, and normal labels but without lesions position indication, released in 2024.

From this survey, we selected CBIS-DDSM and VinDr-Mammo for two complementary reasons. First, they represent the two ends of the image quality spectrum—digitized analog film versus native full-field digital mammography (FFDM)—directly enabling Question (5) about the robustness of findings across varying image quality. Second, both contain a sufficient number of images to train deep learning breast cancer classifiers, provide rigorously validated ground-truth labels (biopsy-confirmed benign/malignant categories in CBIS-DDSM, and BI-RADS assessments independently double-read with arbitration for discordance in VinDr-Mammo), and include lesion-level annotations required for the patch-based pretraining methodology investigated in this work. Among the FFDM datasets surveyed, VinDr-Mammo was preferred over alternatives for the following reasons: RSNA, despite its large size, does not provide lesion position annotations; CMMD contains fewer images than VinDr-Mammo and likewise lacks lesion ROI annotations; and EMBED is only partially available for public download. The remaining datasets were excluded due to limitations such as small size (InBreast, MIAS, IRMA), labels that are neither biopsy-confirmed nor independently arbitrated (KAU-BCMD), labels reflecting masking difficulty rather than diagnostic categories (CSAW-M), or lack of lesion position information (NL-Breast-Screening).

## 2.2 | Related Work

Below, we outline neural network–based CAD systems for single-view mammogram analysis to address the challenge of mammogram classification.

As mentioned earlier, Shen et al. [4] proposed a strategy based on two transfer-learning stages. They achieved an AUC of 0.87 using a specific selection of CBIS-DDSM images for the test set. Other researchers [25] evaluated Shen's method in the original test set and achieved an AUC of 0.75.

Shu et al. [26] proposed two novel pooling techniques to replace the conventional average-pooling and max-pooling layers, using the same dataset. Their method achieves a maximum AUC of 0.838. However, it is unclear whether they employed the original dataset split, since they report using an 85/15% train/test partition, whereas the original dataset is divided 79/21%.

Wei et al. [25], also using the CBIS-DDSM dataset, introduced the use of neural-network morphing in lieu of traditional transfer learning. Using the original train/test split, they report AUCs of 0.7964 for the single model without test-time augmentation (TTA), 0.8187 for the single model with TTA, and 0.8313 for an ensemble of four models with TTA. By contrast, they obtain an AUC of 0.9427 when employing a fixed random train/test split; these results are not directly comparable due to different test splits.

Panceri et al. [27] conducted a study in which they used only 503 craniocaudal (CC) views containing calcification lesions, randomly selected from the 717 CBIS-DDSM images meeting these criteria. Their approach is patch-based: they extract 256 × 256 patches from the selected images, trained a patch classifier, and then apply this classifier to entire test images by dividing them into patches and classifying each one. They aggregate the per-patch probabilities to produce an overall mammography diagnosis.

Bhat et al. [28] employ a technique called AUCReshaping, which aims to modify the ROC curve to improve evaluation within specific sensitivity and specificity ranges. Using the VinDr-Mammo dataset, they evaluated two BI-RADS-based categories, grouping 1, 2, and 3 as the negative class and 4 and 5 as the positive class, achieving an AUC of 0.77. This grouping yields a different class balance than our further experiments.

Shah et al. [7] explored backbone architectures such as AlexNet [29], ResNet [30], DenseNet [31] and EfficientNet [32] using a dataset composed of DDSM [13], RSNA [22] and synthetic images, totaling 6210 mammograms. Their experiments reported an AUC of 94% in both EfficientNet and DenseNet.

On the other hand, multi-view systems aim to improve performance by processing more than one view simultaneously. Approaches vary significantly as below.

Khan et al. [8] introduce a multi-view, patch-based approach using a simplified VGG-style convolutional network to analyze mammographic ROIs from the MIAS and CBIS-DDSM datasets. In three sequential training stages, their model first distinguishes normal from abnormal patches on MIAS with an AUC of 0.934, then classifies CBIS-DDSM patches as masses versus calcifications, AUC of 0.923, and finally differentiates benign from malignant lesions, AUC of 0.769.

Petrini et al. [6] enhanced the results of Shen et al. work [4] by adopting a more advanced convolutional backbone and introducing a two-view classifier. In this design, CC (craniocaudal) and MLO (mediolateral-oblique) feature maps are fused via 2D concatenation and processed jointly by convolutional layers, following the pretraining of a patch-based model and a single-view classifier. Evaluated on the CBIS-DDSM dataset using its original train/test split, the two-view system achieved an AUC of 0.8418 $\pm$ 0.0258.

Chen et al. [33] proposed a multi-view system that uses an EfficientNet-b0 backbone, resized images to 1536 × 768, and processed independently global features and region-level features, passing the later through multi-head attention, from both CC and MLO views, then fuses both for subsequent classification. They trained in a private dataset and tested in CMMD [23] (complete dataset) and InBreast [16] (test division).

Nguyen et al. [9] introduced a two-view classification framework that feeds paired CC and MLO mammograms into parallel ResNet backbones. The extracted feature maps are merged in a fusion block—via averaging, concatenation, and element-wise addition—and subsequently classified by fully connected layers. When evaluated on the CMMD and VinDr-Mammo datasets (with BI-RADS 2 as the negative class and BI-RADS 5 as the positive class), this approach achieved a peak AUC of 0.7486.

Sarker et al. [10] introduced a Swin Transformer Multi-Headed Dynamic Attention module, resized all images to 224 × 224 and 384 × 384 and evaluated on CBIS-DDSM and VinDr-Mammo datasets. This approach, exclusively based on transformers, achieved an AUC of 71.37 using the two views mass mammograms from CBIS-DDSM.

Shah et al. [34] proposed a two-view architecture in which mammographic images were processed through EfficientNet backbones, followed by feature fusion. Using the RSNA dataset, they achieved an accuracy of 0.99, outperforming other comparative methods.

## 2.3 | Methodology

### 2.3.1 | Research Questions and Exploration Strategy

In this research paper, after analyzing multi-view mammogram classification approaches, we formulated the following key questions:

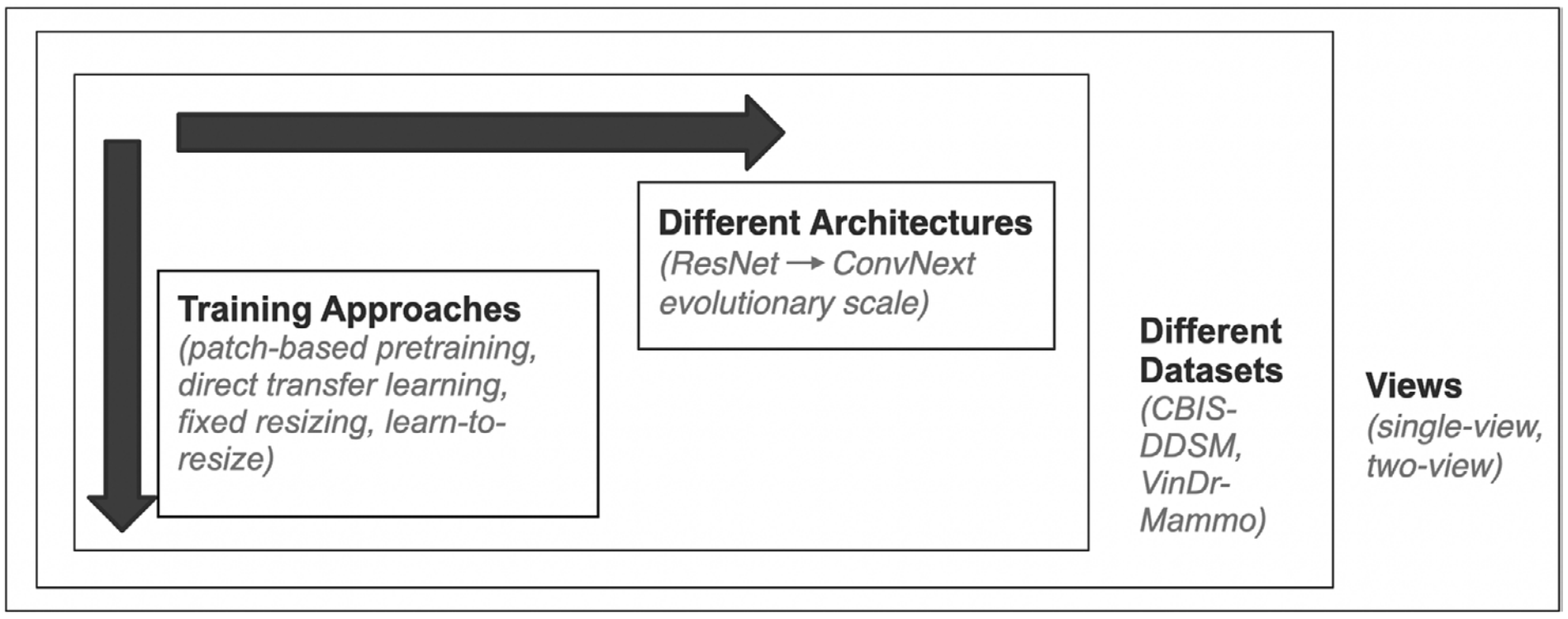

**FIGURE 2** | Exploration methodology.

1. *Are patch classifiers necessary?* Some approaches like Shen et al. [4] and Petrini et al. [6] incorporated a patch classifier as an intermediate step in the transfer learning process. Is this step truly essential?

2. *What is the most suitable backbone for mammogram classification?* Do backbone models that perform well on ImageNet classification also achieve better results on mammograms? Are models trained on ImageNet with a larger dataset (21 or 22k categories) superior to those trained on the smaller dataset (1k categories)? Do models with higher input image resolutions yield better performance for mammogram classification?

3. *Can mammogram resolution be reduced without significantly affecting classification performance?* Lower-resolution mammograms reduce computational resource requirements for processing. One promising approach is the “learn-to-resize” technique [35], which adaptively optimizes resizing for specific tasks. How effective is this method when applied to mammogram classification?

4. *Does the two-view classifier provide a significant performance improvement over the single-view classifier?* A two-view classifier cannot be directly compared to a single-view classifier, as the former produces $n$ results for a dataset with $2n$ images, while the latter generates $2n$ results. To ensure a fair comparison, we combined the results of the single-view classifier applied to each mammographic view (CC and MLO) using both average and maximum operations, reducing the output to $n$ results.

5. *How does image quality affect the answers to the above questions?* These questions can be explored using both low-quality analog mammograms (such as CBIS-DDSM [17]) and high-quality digital mammograms (such as VinDr-Mammo [20]). Would the answers differ depending on the quality of the mammographic images?

By investigating these questions, we aim to refine best practices in machine learning-based mammogram classification, optimizing both accuracy and computational efficiency. In doing so, we developed models that surpass previous results for both single-view and two-view classifiers on the CBIS-DDSM and VinDr-Mammo public datasets. To select the base model architectures, inspired by [36], we will follow an evolutionary scale starting from ResNet [30] until ConvNeXt [37], using open source implementations from Wightman [38] and from PyTorch [39] and will evaluate different training approaches, like patch-based pretrain, resizing and others, as summarized in Figure 2.

**TABLE 1** | Division of CBIS-DDSM dataset into training, validation, and test sets.

| Type | Training | Validation | Test |
|---|---|---|---|
| Mammograms | 2212 | 246 | 645 |

### 2.3.2 | Datasets and Splits

First, we will conduct all experiments using the CBIS-DDSM dataset, which consists of low-quality digitized analog mammograms. All images were resized to $1152 \times 896$ pixels before processing. The CBIS-DDSM dataset was used in its entirety. We used the original test set as test data in the present study, to allow comparisons with other works that use the original test set as [6, 25], and the total number images are show in Table 1.

Then, we will repeat the experiments using the VinDr-Mammo dataset [20], which consists entirely of Full-Field Digital Mammographies (FFDMs) to answer the Question (5). Unlike CBIS-DDSM, this dataset does not include biopsy-confirmed benign or malignant labels. Instead, it provides Bi-RADS (Breast Imaging Reporting and Data System) and other annotations, uniformly distributed across 4000 training exams and 1000 test exams.

The distribution of Bi-RADS categories is: 1 (67.03%), 2 (23.38%), 3 (4.65%), 4 (3.81%), and 5 (1.13%). To evaluate the performance of full-image classifiers, we grouped the Bi-RADS categories into two broader classes: “Normal” for views classified as Bi-RADS 1 and 2, and “Abnormal” for views classified as Bi-RADS 3, 4, and 5. This grouping was based on the presence of lesion annotations in the latter categories. As a result, the “Abnormal” class represents approximately 10% of the dataset. We assessed the target task of categorizing mammograms into these two classes (Table 2).

To construct the patch dataset, outlined in Table 3, we utilized the lesion annotations provided in the VinDr-Mammo dataset. As a fourth category, we selected background samples in the abnormal images excluding lesion areas.

**TABLE 2** | Number of VinDr-Mammo images used in this experiment.

| Type | Training | Validation | Test |
|---|---|---|---|
| Mammograms | 14 394 | 1604 | 4000 |
| Abnormal | 1380 | 154 | 384 |
| Normal | 13 014 | 1450 | 3616 |

**TABLE 3** | Number of VinDr-Mammo patches used in this work.

| Type | Training | Validation | Test |
|---|---|---|---|
| Bi-Rads 3 | 5820 | 710 | 1680 |
| Bi-Rads 4 | 6320 | 680 | 1810 |
| Bi-Rads 5 | 2590 | 400 | 810 |
| Background | 13 567 | 1647 | 3963 |

### 2.3.3 | Training Protocol

Each network was trained three times under identical hyperparameters and data splits, with only the random seed varied between runs. This seed controls weight initialization, training-data shuffling order, and stochastic data augmentation sampling, so the three resulting models reflect the inherent variability of stochastic gradient-based training rather than any deliberate architectural or data difference. From each training round, we selected the checkpoint with the highest performance on the validation set and evaluated it on the test set, recording this result as "Best Test."

### 2.3.4 | Statistical Analysis

**2.3.4.1 | Performance Metrics.** For patch classifiers, we report classification accuracy. For full-image classifiers (single-view and multi-view), we report the Area Under the Receiver Operating Characteristic Curve (AUC), which is well-suited to imbalanced binary classification problems such as normal-versus-abnormal mammogram categorization. The "Test Mean" reported in subsequent tables is the average test-set performance across the three independently trained models described in Section 2.3.3, with the corresponding standard deviation providing a measure of run-to-run variability.

The standard error of the AUC for "Best Test" was computed using the Hanley and McNeil method [40], which estimates the standard error from the AUC value and the numbers of positive and negative cases in the test set.

**2.3.4.2 | Hypothesis Testing Between Two AUCs.** To compare two AUCs obtained on the same test set, we used two complementary statistical procedures. The first is a normal-approximation z-test based on the Hanley and McNeil formula [41] for the standard error of a difference in correlated AUCs. Specifically, given two AUCs ($AUC_1$, $AUC_2$) and their standard errors ($SE_1$, $SE_2$), we compute:

$$z = \frac{AUC_1 - AUC_2}{SE_{\text{diff}}}, SE_{\text{diff}} = \sqrt{SE_1^2 + SE_2^2 - 2 \cdot r \cdot SE_1 \cdot SE_2}$$

where $r$ denotes the correlation between the two AUCs, we assume a conservative Kendall-style value of $r = 0.5$. The one-sided $p$ value is then obtained from the standard normal distribution. We refer to this procedure as the "hypothesis test" throughout the paper.

The second procedure is the DeLong test [42], a nonparametric method that directly estimates the covariance structure of two correlated AUCs from the paired case-level predictions, without assuming a specific value for $r$. We used the fast one-tailed implementation of Sun and Xu [43], with the open-source implementation from [44]. When paired case-level predictions were available, we preferred the DeLong test for its statistical rigor. When only summary AUCs and their standard errors were available, we used the Hanley-McNeil-based hypothesis test.

**2.3.4.3 | Significance Level and Reporting.** Throughout the paper, we report exact $p$ values rather than fixed significance thresholds, allowing the reader to assess the strength of each statistical claim. When required by convention, we use $\alpha = 0.05$ as the nominal significance threshold. All $p$ values are one-sided, testing the directional hypothesis that the first-listed AUC exceeds the second.

**2.3.4.4 | Correlation Analysis.** To assess whether backbones that perform better on natural-image classification also perform better on mammogram classification, we computed the Pearson correlation coefficient $r$ between the top-1 accuracy of each backbone on ImageNet-1k (as reported by its original authors) and the AUC of the corresponding mammogram classifier under a given approach. The Pearson coefficient measures the strength and direction of a linear association between two variables and takes values in $[-1, 1]$, with $r = 1$ indicating perfect positive linear correlation, $r = 0$ indicating no linear association, and $r = -1$ indicating perfect negative correlation. Values of $|r|$ between 0.3 and 0.7 are conventionally interpreted as moderate correlation, and $|r| \geq 0.7$ as strong correlation.

### 2.3.5 | Patch Classifier and Single-View Classifiers

In this section, we aim to address questions (1) and (2). To investigate these, we perform two sets of experiments using different approaches to initialize the weights of single-view classifiers:

1. Patch-based classifier (PBC): Train a patch classifier and leverage transfer learning from its weights.
2. Direct classifier (DC): Apply transfer learning directly from weights pretrained on ImageNet.

The use of patches as an intermediate training step is motivated by domain-specific transfer learning rather than by GPU memory constraints. Mammograms differ substantially from natural images in their statistical properties: they are single-channel, low-contrast, and contain subtle, localized lesions against a largely homogeneous background. As a result, weights pretrained on ImageNet [5], which encode features tuned to natural-image textures and object categories, may transfer suboptimally to mammogram classification. The patch classifier addresses this gap: by training a CNN to distinguish lesion categories (e.g., benign mass, malignant calcification, background) directly on $224 \times 224$ mammogram crops, the network learns feature representations specifically tuned to mammographic pathology at the local scale. These domain-adapted weights then serve as a stronger initialization for the whole-image classifier than ImageNet weights would provide. Importantly, the patches in this work are extracted from already-resized $1152 \times 896$ mammograms rather than from native acquisition resolution, so memory considerations play no role in this design choice, as the whole-image classifier operates at the same $1152 \times 896$ resolution.

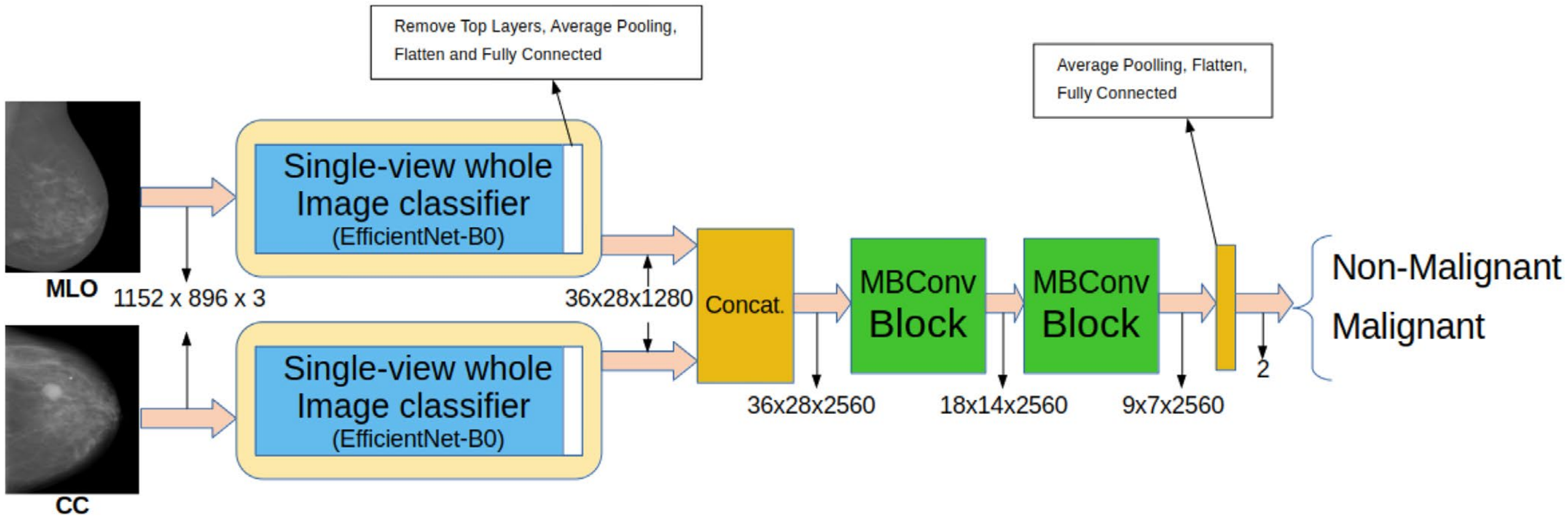

**FIGURE 3** | Two-view classifier with EfficientNet-B0 backbone.

The CBIS-DDSM provides annotated ROIs in the form of binary masks—one per lesion, with the same dimensions as the original image—in which the lesion boundaries are segmented. When overlaid on the original mammogram, these masks delineate the borders of a mass or the extent of a calcification region. Considering it, we extracted $224 \times 224$ pixel fragments (patches) from these regions, dividing them in five categories: background, benign calcification, malignant calcification, benign mass, or malignant mass. For every lesion, we generated 10 patches by randomly shifting the center of mass within a ±10% range in various directions. Additionally, we extracted 10 background patches per image to ensure balanced representation.

The patch classifier is initialized with weights pretrained on ImageNet and trained using the Adam optimizer, starting with an initial learning rate of $2 \times 10^{-5}$. The learning rate follows a “warm-up and cyclic cosine” schedule, configured with a period of 3, a maximum learning rate delta of $2 \times 10^{-4}$, and a warm-up phase lasting 4 epochs.

**2.3.5.1 | Patch Based Classifier (PBC).** We trained the models to classify single-view mammograms by leveraging the patch classifier weights obtained in the previous paragraph and taking advantage of the fully convolutional network architecture, which enables the classification of images with varying sizes. To adapt the architecture for this task, we added two EfficientNet MBConv blocks (with stride = 2) as the top layers, followed by a fully connected layer, as the single image classifier in [6].

**2.3.5.2 | Direct Classifier (DC).** We trained the single-view classifiers, now as pure backbones, using transfer learning directly from base models pretrained on ImageNet, bypassing the patch classifier.

**2.3.5.3 | VinDr-Mammo Patches.** To answer Questions (1) and (2) for VinDr-Mammo dataset, we prepared the patches similar to CBIS-DDSM method mentioned above, considering four categories: background and the lesions with Bi-Rads 3, 4 and 5.

### 2.3.6 | Resolution Reduction

To address Question (3), we tested reducing the input image size by half, resulting in dimensions of $576 \times 448$. We conducted two experiments using different downsampling methods. The first experiment employed conventional interpolation, while the second utilized a machine learning-based downsampling technique called “Learn to Resize” [35]. If this technique demonstrates superior performance compared to conventional resampling, it could become the preferred method for reducing high-resolution mammograms to a resolution compatible with current processing devices.

For the fisrt experiment, we downscaled the original $1152 \times 892$ images to $576 \times 448$ using the INTER_AREA interpolation method from the OpenCV library.

For the second experiment, we used “Learn to Resize,” which integrates bilinear resizing with convolutional layers, enabling the model to optimize resizing for improved classification performance.

### 2.3.7 | Multi-View Classifier

We assembled the network as in Figure 3 making another transfer learning with the weights obtained from the best single-view models.

The results of two-view and single-view classifiers cannot be directly compared, as they evaluate different units: a two-view classifier produces n predictions for a dataset with 2n images, while a single-view classifier produces 2n predictions. To address this, we propose a per-exam aggregation protocol, as mentioned in Section 2.3.1, item (4). The resulting in per-exam single-view predictions can then be compared directly against the n per-exam two-view predictions on identical test units, enabling a valid significance test.

#### 2.3.8 | Cross-Dataset Evaluation Protocol

To address Question (5), the methodologies described in Sections 2.3.3–2.3.7 were replicated on both datasets under identical experimental protocols. In both cases, we used each dataset's original train/test split to avoid any selection effects that custom splits might introduce. The only differences between the two protocols were: (i) the patch label categories (five for CBIS-DDSM versus four for VinDr-Mammo, reflecting the available annotations); and (ii) the binary classification target (benign/malignant for CBIS-DDSM versus normal/abnormal grouped from BI-RADS labels for VinDr-Mammo). This parallel design enables direct comparison of how each design choice behaves across image quality regimes.

## 3 | Results and Discussion

In this section we apply the methodology in both public datasets, firstly in CBIS-DDSM and later in VinDr-Mammo. The "ImageNet input size" column, in results tables below, indicates the input resolution at which each backbone was pretrained on ImageNet (not the mammogram resolution, which is fixed at 1152 × 896, unless otherwise noted).

### 3.1 | Single View Classifiers in CBIS-DDSM

#### 3.1.1 | Patch Classifier and Base-Model

The results, summarized in Table 4, demonstrate that among the evaluated base models, ConvNeXt-Base achieves the highest performance, with an accuracy of 0.7918.

**TABLE 4** | Accuracy of patch classifiers. Models pretrained on 21k or -22k are explicitly marked, while those trained on ImageNet-1k are unmarked.

| Model | Test mean ± standard deviation | Best test |
|---|---|---|
| ViT-Base-CLIP-Laion-32 | 0.6464 ± 0.0109 | 0.6487 |
| MNASNet-1.0 | 0.6901 ± 0.0269 | 0.7152 |
| MobileNetV2 | 0.7391 ± 0.0039 | 0.7369 |
| EfficientNetV2-S (21k) | 0.7524 ± 0.0056 | 0.7460 |
| MobileNetV3-Large | 0.7357 ± 0.0100 | 0.7462 |
| EfficientNet-B1 | 0.7585 ± 0.0090 | 0.7495 |
| DenseNet169 | 0.7594 ± 0.0053 | 0.7540 |
| ResNet-18 | 0.7508 ± 0.0075 | 0.7579 |
| ResNeXt-50-32x4d | 0.7580 ± 0.0051 | 0.7585 |
| DenseNet201 | 0.7617 ± 0.0014 | 0.7604 |
| EfficientNet-B0 | 0.7585 ± 0.0035 | 0.7609 |
| EfficientNet-B2 | 0.7550 ± 0.0077 | 0.7637 |
| EfficientNetV2-M | 0.7501 ± 0.0129 | 0.7641 |
| ResNet-101 | 0.7574 ± 0.0087 | 0.7644 |
| ResNet-50 | 0.7550 ± 0.0087 | 0.7650 |
| EfficientNet-B4 | 0.7639 ± 0.0057 | 0.7666 |
| ConvNeXt-Base (22k) | 0.7740 ± 0.0063 | 0.7680 |
| DenseNet121 | 0.7685 ± 0.0119 | 0.7691 |
| EfficientNetV2-S | 0.7643 ± 0.0111 | 0.7695 |
| SwinV2-Base-In22k-FT-1k (22k) | 0.7725 ± 0.0048 | 0.7695 |
| EfficientNet-B3 | 0.7661 ± 0.0050 | 0.7700 |
| ConvNeXt-Tiny | 0.7711 ± 0.0051 | 0.7721 |
| ConvNeXt-Small | 0.7791 ± 0.0081 | 0.7784 |
| ConvNeXt-Base | 0.7831 ± 0.0082 | **0.7918** |

#### 3.1.2 | Patch Based Classifier (PBC)

The resulting AUCs are presented in Table 5. Notably, we observed that the best-performing base models for patch classification did not consistently perform well for full mammogram classification. The best model, EfficientNet-B3, achieved an AUC of 0.8325±0.171, however, executing the hypothesis test from page 6, we obtained a $p$ value of 0.1588, which shows no significant superiority over the second best classifier, EfficientNetV2-M (AUC = 0.8150 ± 0.0179).

#### 3.1.3 | Direct Classifier (DC)

The resulting AUCs are summarized in Table 6. Consistent with previous findings, EfficientNet-B3 achieved the best performance among the evaluated base models, with an AUC of 0.8313 ± 0.0172. However, it presented significant similarity with the second best classifier, EfficientNetV2-S (21k) (AUC = 0.8301 ± 0.0172), with a $p$ value of 0.4722.

#### 3.1.4 | Conclusions About Single View Classifiers

As shown in Table 7, PBC and DC yield comparable performance when implemented with the EfficientNet-B3 backbone. Executing the hypothesis test, described on page 6, we obtained a $p$ value of 0.4721, suggesting that a patch-based classifier (PBC) does not offer a significant advantage over directly

**TABLE 5** | AUCs obtained by single-view patch-based classifiers (PBC).

| Model | ImageNet input size | Test mean ± standard deviation | Best test | PBC/DC |
|---|---|---|---|---|
| MobileNetV3-Large | 224 | 0.7426 ± 0.0065 | 0.7353 ± 0.0205 | DC |
| MobileNetV2 | 224 | 0.7504 ± 0.0315 | 0.7611 ± 0.0198 | DC |
| ResNet-50 | 224 | 0.7541 ± 0.0207 | 0.7616 ± 0.0197 | DC |
| ResNeXt-50-32x4d | 224 | 0.7705 ± 0.0068 | 0.7627 ± 0.0197 | PBC |
| DenseNet121 | 224 | 0.7804 ± 0.0161 | 0.7633 ± 0.0197 | PBC |
| MNASNet-1.0 | 224 | 0.7627 ± 0.0066 | 0.7643 ± 0.0197 | PBC |
| EfficientNet-B4 | 384 | 0.7866 ± 0.0040 | 0.7814 ± 0.0191 | DC |
| ConvNeXt-Tiny | 224 | 0.7969 ± 0.0102 | 0.7850 ± 0.0190 | PBC |
| ResNet-18 | 224 | 0.7749 ± 0.0129 | 0.7858 ± 0.0189 | PBC |
| EfficientNetV2-S | 300 | 0.8034 ± 0.0117 | 0.7886 ± 0.0189 | DC |
| DenseNet201 | 224 | 0.7884 ± 0.0129 | 0.7891 ± 0.0188 | PBC |
| ResNet-101 | 224 | 0.7791 ± 0.0093 | 0.7892 ± 0.0188 | PBC |
| EfficientNet-B0 | 224 | 0.7933 ± 0.0096 | 0.7977 ± 0.0185 | DC |
| ConvNeXt-Small | 224 | 0.8010 ± 0.0033 | 0.7984 ± 0.0185 | PBC |
| ConvNeXt-Base | 224 | 0.7991 ± 0.0045 | 0.8019 ± 0.0184 | PBC |
| ConvNeXt-Base (22k) | 224 | 0.8137 ± 0.0108 | 0.8027 ± 0.0183 | PBC |
| EfficientNet-B2 | 260 | 0.8054 ± 0.0128 | 0.8120 ± 0.0180 | PBC |
| DenseNet169 | 224 | 0.7910 ± 0.0106 | 0.8031 ± 0.0183 | PBC |
| EfficientNet-B1 | 240 | 0.7884 ± 0.0120 | 0.8054 ± 0.0182 | DC |
| EfficientNetV2-S (21k) | 300 | 0.8089 ± 0.0052 | 0.8114 ± 0.0180 | DC |
| EfficientNetV2-M | 384 | 0.7931 ± 0.0158 | 0.8150 ± 0.0179 | PBC |
| EfficientNet-B3 | 300 | 0.8114 ± 0.0153 | **0.8325 ± 0.0171** | PBC |

*Note:* PBC/DC indicates which approach generates the highest AUC in the best test, comparing with Table 6.

utilizing ImageNet-pretrained weights (DC). Moreover, when Test Time Augmentation (TTA) was applied, no statistical significance between approaches was found either, with $p$ value = 0.4649. Additionally, while 14 base models performed better under the PBC approach, eight showed improved results with DC. These findings further reinforce the conclusion that there is no clear benefit to using PBC over DC.

Thus, we can answer Question (1): since PBC is more complex and time-consuming to train than DC, we conclude that pretraining on patches is not a suitable strategy for building classifiers for lower-quality mammography datasets such as CBIS-DDSM.

### 3.1.5 | Conclusions About Base-Models

In this section, we analyze Tables 5 and 6 to address the Question (2), concluding:

- For PBC, EfficientNet-B3 is significantly superior to DenseNet169, the best-performing network outside the EfficientNet family, with $p = 0.0491$, *suggesting that meaningful performance differences exist only within the EfficientNet family*. For DC, EfficientNet-B3 shows no statistically significant advantage over MobileNetV3-Large—the best non-EfficientNet model in that setting—with $p = 0.1525$.
- *ImageNet-22k or -21k initializations presented heterogeneous results compared to ImageNet-1k*. For ConvNeXt, the difference was not significant for PBC ($p = 0.4826$) but was significant for DC ($p = 0.0030$). For EfficientNetV2-S, neither PBC ($p = 0.1084$) nor DC ($p = 0.4199$) showed a statistically significant difference.
- *Larger pretraining image sizes* (*300 × 300 or greater*) were associated with better performance. For PBC, the best model pretrained on 300 × 300 ImageNet input size, presented some statistical significance, over the best 224 × 224 model ($p = 0.0631$). For DC, it was not significantly superior, with $p = 0.1515$.
- *Correlation analysis between base model performance and classifier performance* is illustrated in Figure 4. We computed the Pearson correlation coefficient ($r$) between the AUCs of PBCs and the top-1 accuracy of their base models on ImageNet, yielding $r = 0.65$, indicating a moderate positive correlation. This suggests that PBCs generally benefit from stronger base models. In contrast, DC classifiers

**TABLE 6** | AUCs obtained by single-view direct classifiers (DC).

| Model | ImageNet input size | Test mean ± standard deviation | Best test | PBC/DC |
|---|---|---|---|---|
| DenseNet169 | 224 | 0.6784 ± 0.0184 | 0.6524 ± 0.0222 | PBC |
| DenseNet121 | 224 | 0.6697 ± 0.0057 | 0.6696 ± 0.0219 | PBC |
| ConvNeXt-Base | 224 | 0.6817 ± 0.0178 | 0.6701 ± 0.0219 | PBC |
| ConvNeXt-Tiny | 224 | 0.6889 ± 0.0123 | 0.6791 ± 0.0218 | PBC |
| ConvNeXt-Small | 224 | 0.6674 ± 0.0152 | 0.6802 ± 0.0217 | PBC |
| DenseNet201 | 224 | 0.6878 ± 0.0078 | 0.6812 ± 0.0217 | PBC |
| ResNeXt-50-32x4d | 224 | 0.6876 ± 0.0054 | 0.6909 ± 0.0215 | PBC |
| ResNet-18 | 224 | 0.6883 ± 0.0073 | 0.6935 ± 0.0215 | PBC |
| ConvNeXt-Base (22k) | 224 | 0.7019 ± 0.0262 | 0.7289 ± 0.0207 | PBC |
| MNASNet-1.0 | 224 | 0.7633 ± 0.0147 | 0.7427 ± 0.0203 | PBC |
| MobileNetV2 | 224 | 0.7584 ± 0.0251 | 0.7659 ± 0.0196 | DC |
| ResNet-101 | 224 | 0.7698 ± 0.0092 | 0.7702 ± 0.0195 | PBC |
| EfficientNet-B4 | 384 | 0.8023 ± 0.0064 | 0.7933 ± 0.0187 | DC |
| ResNet-50 | 224 | 0.7293 ± 0.0598 | 0.7993 ± 0.0185 | DC |
| EfficientNetV2-M | 384 | 0.7605 ± 0.0376 | 0.8024 ± 0.0183 | PBC |
| EfficientNet-B2 | 260 | 0.8058 ± 0.0071 | 0.8078 ± 0.0181 | PBC |
| EfficientNet-B1 | 240 | 0.8023 ± 0.0156 | 0.8117 ± 0.0180 | DC |
| EfficientNet-B0 | 224 | 0.7691 ± 0.0525 | 0.8127 ± 0.0179 | DC |
| MobileNetV3-Large | 224 | 0.7792 ± 0.0247 | 0.8132 ± 0.0179 | DC |
| EfficientNetV2-S | 300 | 0.8100 ± 0.0251 | 0.8266 ± 0.0174 | DC |
| EfficientNetV2-S (21k) | 300 | 0.8098 ± 0.0162 | 0.8301 ± 0.0172 | DC |
| EfficientNet-B3 | 300 | 0.8087 ± 0.0177 | **0.8313 ± 0.0172** | PBC |

*Note:* PBC/DC indicates which approach generates the highest AUC in the best test, comparing with Table 5.

**TABLE 7** | AUCs of the best-performing single-view classifiers on the CBIS-DDSM dataset.

| Classifier | Network | Result | TTA |
|---|---|---|---|
| Reference [6] | EfficientNet-B0 | 0.8033 ± 0.0183 | 0.8153 ± 0.0178 |
| PBC | EfficientNet-B3 | **0.8325 ± 0.0171** | 0.8343 ± 0.0170 |
| DC | EfficientNet-B3 | 0.8313 ± 0.0172 | **0.8358 ± 0.0170** |

showed a weak correlation ($r = 0.21$), implying that their performance is less dependent on the base model's ImageNet accuracy.

#### 3.1.6 | Comparison With Previous Work

The AUCs achieved by our patch-based classifier (PBC—0.8325) and direct classifier (DC—0.8313) are significantly higher than the best result reported in [6] (0.8033), see in Table 7. Using the hypothesis test described on page 6, we compared the AUC of our PBC model with the previous work, obtaining a $p$ value of 0.0499. Similarly, the comparison for our DC model yielded a $p$ value of 0.0575. These results demonstrate that the improvements in our work are substantial, with some statistical significance. We also achieved better results than Wei et al. [25] (81.87%), however it is not possible to calculate statistical significance as there are no standard errors available in this work. We attribute these gains to the use of modern pretraining strategies for ImageNet-based models. For instance, He et al. [45] improved the top-1 accuracy of ResNet50 from 75.3% to 79.29%, with 2.13% of this improvement directly attributable to advanced techniques such as cosine learning rate decay, label smoothing, and Mixup. Further enhancements by Wightman et al. [46] increased this accuracy to 80.4% through the incorporation of the LAMB optimizer and CutMix augmentation.

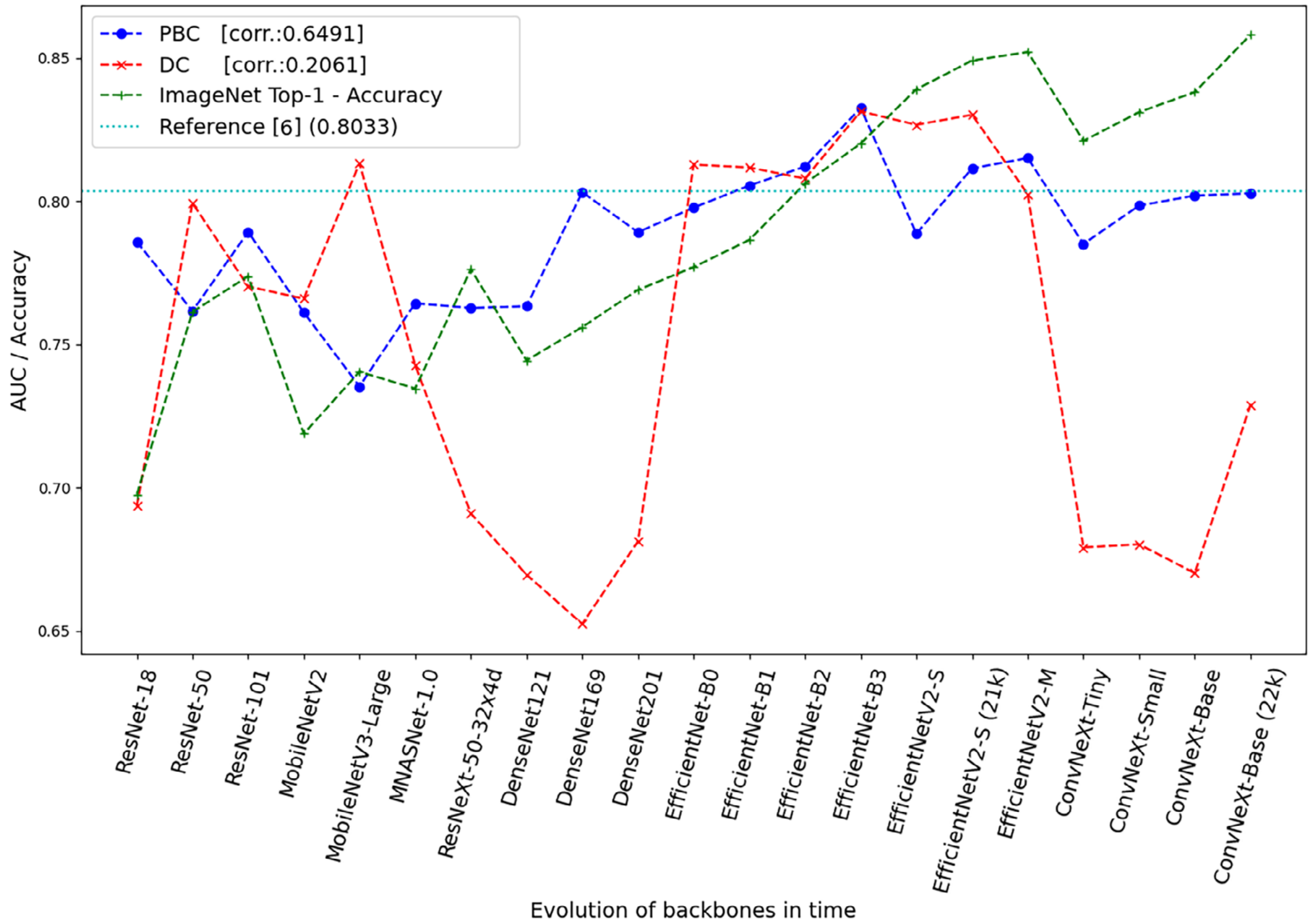

**FIGURE 4** | AUC values for all single-view PBC and DC classifiers on the CBIS-DDSM dataset, along with the accuracies of their backbone models on ImageNet classification, and the corresponding Pearson correlation coefficients.

In the Test Time Augmentation (TTA) experiments, the direct classifier (DC) in this study achieved an AUC of 0.8343, compared to the AUC of 0.8153 reported in [6] (not statistically superior, with a $p$ value of 0.1384) and Wei et al. [25] (83.13%).

## 3.2 | Classifiers With Resized Images in CBIS-DDSM

### 3.2.1 | Fixed Resizing Classifier (FRC)

Transfer learning was then performed directly from the ImageNet pretrained weights. The results, presented in Table 8, show that the highest AUC achieved was $0.8167 \pm 0.0178$ with a $p$ value of 0.2565 comparing with the second best, which shows that is not a significant superiority. This performance is notably lower than the results obtained without downsampling: 0.8313 $\pm$ 0.0172 for direct classification (DC), however not significantly, with a $p$ value of 0.2020.

### 3.2.2 | Learn-to-Resize Classifier (LRC)

As shown in Table 9, the highest AUC achieved using this method was $0.7958 \pm 0.0186$ (not significantly different from next result, with a $p$ value of 0.3582).

### 3.2.3 | Conclusions on Resizing

Table 10 summarizes the highest AUCs obtained with and without resizing. The results show that the "Learn to Resize" technique (LRC) underperformed fixed resizing (FRC), though this difference was not statistically significant ($p = 0.1257$). Given its greater computational complexity, LRC offers no practical advantage over FRC, suggesting it may not be well-suited for mammograms despite its effectiveness on natural images. Additionally, downscaling the input images led to a decrease in AUC, though this reduction did not reach statistical significance.

Hypothesis test described in page 6 reveals that patch-based classification (PBC) is significantly superior to LRC, with a $p$ value of 0.0202. However, there is limited statistical evidence to conclude that PBC outperforms FRC, as the $p$ value for this comparison is 0.1828.

In our tests, larger images yielded better results; however, a different study [47] demonstrated that higher resolutions ($2304 \times 1792$) did not improve performance.

From Tables 8 and 9, we observed that the best-performing networks using FRC and LRC utilized backbones pretrained on $224 \times 224$ ImageNet inputs, however, their superiority over other pretrain image sizes carries low statistical significance, as indicated by $p$ values of 0.2565 and 0.3582 for FRC and LRC, respectively.

## 3.3 | Two-View Classifiers in CBIS-DDSM

In this section, we aim to address the top Question (4):

By assembling two-view networks using the best single-view networks weights, as in Figure 3, we obtained the AUCs described in Table 11.

**TABLE 8** | AUCs of single-view direct classifiers using fixed resizing (FRC).

| Model | ImageNet input size | Test mean ± standard deviation | Best test |
|---|---|---|---|
| MNASNet-1.0 | 224 | 0.5643 ± 0.0267 | 0.5860 ± 0.0230 |
| MobileNetV2 | 224 | 0.6858 ± 0.0169 | 0.6972 ± 0.0214 |
| ConvNeXt-Small | 224 | 0.7084 ± 0.0240 | 0.7117 ± 0.0211 |
| ConvNeXt-Tiny | 224 | 0.7118 ± 0.0330 | 0.7214 ± 0.0209 |
| ResNet-18 | 224 | 0.7433 ± 0.0274 | 0.7536 ± 0.0200 |
| EfficientNet-B2 | 260 | 0.7622 ± 0.0143 | 0.7656 ± 0.0196 |
| EfficientNet-B1 | 240 | 0.7644 ± 0.0077 | 0.7680 ± 0.0195 |
| DenseNet121 | 224 | 0.7896 ± 0.0106 | 0.7751 ± 0.0193 |
| MobileNetV3-Large | 224 | 0.7811 ± 0.0131 | 0.7753 ± 0.0193 |
| EfficientNetV2-S (21k) | 300 | 0.7808 ± 0.0071 | 0.7761 ± 0.0193 |
| EfficientNetV2-M | 384 | 0.7723 ± 0.0269 | 0.7832 ± 0.0190 |
| ResNeXt-50-32x4d | 224 | 0.7817 ± 0.0137 | 0.7881 ± 0.0189 |
| EfficientNet-B4 | 384 | 0.7847 ± 0.0053 | 0.7890 ± 0.0188 |
| DenseNet201 | 224 | 0.7813 ± 0.0107 | 0.7943 ± 0.0186 |
| EfficientNet-B0 | 224 | 0.7686 ± 0.0242 | 0.7956 ± 0.0186 |
| ConvNeXt-Base (22k) | 224 | 0.7951 ± 0.0057 | 0.7958 ± 0.0186 |
| DenseNet169 | 224 | 0.7795 ± 0.0124 | 0.7968 ± 0.0186 |
| EfficientNet-B3 | 300 | 0.8010 ± 0.0175 | 0.7972 ± 0.0185 |
| ResNet-101 | 224 | 0.7876 ± 0.0207 | 0.7981 ± 0.0185 |
| ConvNeXt-Base | 224 | 0.7423 ± 0.0440 | 0.8009 ± 0.0184 |
| EfficientNetV2-S | 300 | 0.7901 ± 0.0102 | 0.8049 ± 0.0183 |
| ResNet-50 | 224 | 0.7756 ± 0.0300 | **0.8167 ± 0.0178** |

The best two-view classifier was achieved using the direct classification (DC) approach with the EfficientNet-B3 base model, yielding an AUC of 0.8643. This represents the highest AUC recorded for classifying two-view exams in the CBIS-DDSM dataset under its original training/testing split. This result surpasses the AUC of 0.8418 reported in [6] using EfficientNet-B0, with some statistical evidence of superiority ($p$ value = 0.0821, using the hypothesis test from page 6). We attribute this improvement to the use of a base model trained with more advanced and modern techniques, as discussed earlier. Using Test Time Augmentation (TTA), we achieved an AUC of 0.8658, compared to the correspondent AUC of 0.8483 reported in [6]. The hypothesis test yielded a $p$ value of 0.1366, indicating weak statistical evidence that our result is superior to the previous.

### 3.3.1 | Average or Maximum Operation

We compared the best two-view result from Table 11 with the result obtained by making inferences independently for each view (CC and MLO), applying the per-exam aggregation protocol from Section 2.3.7.

We conducted the DeLong test to compare the two-view model with the mean and maximum operations (Table 12). The resulting $p$ values were 0.0280 and 0.0286, respectively.

We can now address Question (4): classifying two views simultaneously is statistically superior to classifying individual views and combining their results using mean or maximum operations. This improvement likely stems from the fact that concatenated views provide additional information, such as the spatial locations of lesions, which is not captured by simply aggregating the outputs of separate views.

## 3.4 | Single View Classifiers in VinDr-Mammo

### 3.4.1 | Patch Classifier and Base-Model

The backbone model with the best performance was ConvNeXt_Base_22k with an accuracy of 0.7052 (Table 13).

**TABLE 9** | AUCs of single-view direct classifiers with learned resizing (LRC).

| Model | ImageNet input size | Test mean ± standard deviation | Best test |
|---|---|---|---|
| DenseNet121 | 224 | 0.6712 ± 0.0208 | 0.6518 ± 0.0222 |
| ResNeXt-50-32x4d | 224 | 0.6735 ± 0.0042 | 0.6682 ± 0.0220 |
| ConvNeXt-Small | 224 | 0.6734 ± 0.0080 | 0.6709 ± 0.0219 |
| DenseNet169 | 224 | 0.6718 ± 0.0178 | 0.6850 ± 0.0216 |
| DenseNet201 | 224 | 0.6848 ± 0.0018 | 0.6855 ± 0.0216 |
| ResNet-18 | 224 | 0.6847 ± 0.0030 | 0.6884 ± 0.0216 |
| ConvNeXt-Base | 224 | 0.6791 ± 0.0076 | 0.6892 ± 0.0216 |
| EfficientNet-B4 | 384 | 0.7501 ± 0.0274 | 0.7113 ± 0.0211 |
| EfficientNet-B0 | 224 | 0.7494 ± 0.0228 | 0.7174 ± 0.0209 |
| ResNet-50 | 224 | 0.7188 ± 0.0082 | 0.7272 ± 0.0207 |
| MobileNetV2 | 224 | 0.7096 ± 0.0269 | 0.7297 ± 0.0206 |
| ResNet-101 | 224 | 0.7230 ± 0.0226 | 0.7532 ± 0.0200 |
| EfficientNet-B1 | 240 | 0.7477 ± 0.0063 | 0.7547 ± 0.0200 |
| EfficientNet-B3 | 300 | 0.7613 ± 0.0094 | 0.7564 ± 0.0199 |
| ConvNeXt-Tiny | 224 | 0.7200 ± 0.0273 | 0.7574 ± 0.0199 |
| MNASNet-1.0 | 224 | 0.6979 ± 0.0464 | 0.7626 ± 0.0197 |
| EfficientNet-B2 | 260 | 0.7497 ± 0.0345 | 0.7690 ± 0.0195 |
| ConvNeXt-Base (22k) | 224 | 0.7399 ± 0.0393 | 0.7707 ± 0.0195 |
| EfficientNetV2-M | 384 | 0.7335 ± 0.0297 | 0.7749 ± 0.0193 |
| EfficientNetV2-S (21k) | 300 | 0.7719 ± 0.0122 | 0.7779 ± 0.0192 |
| EfficientNetV2-S | 300 | 0.7475 ± 0.0342 | 0.7890 ± 0.0188 |
| MobileNetV3-Large | 224 | 0.7724 ± 0.0194 | **0.7958 ± 0.0186** |

**TABLE 10** | AUCs of single view classifiers with and without resizing.

| Classifier | Best network | Result |
|---|---|---|
| FRC (with resizing) | ResNet-50 | 0.8167 ± 0.0178 |
| LRC (with resizing) | MobileNetV3_Large | 0.7958 ± 0.0186 |
| PBC (without resizing) | EfficientNet-B3 | **0.8325 ± 0.0171** |
| DC (without resizing) | EfficientNet-B3 | 0.8313 ± 0.0172 |

**TABLE 11** | AUCs of the top-performing two-view classifiers, built using the best single-view classifiers.

| Model | Test mean ± standard deviation | Best test | TTA |
|---|---|---|---|
| EfficientNet-B3 (PBC) | 0.8523 ± 0.0073 | 0.8468 ± 0.0254 | 0.8490 ± 0.0253 |
| EfficientNet-B3 (DC) | 0.8605 ± 0.0028 | **0.8643 ± 0.0241** | **0.8658 ± 0.0239** |

**TABLE 12** | AUC of the two-view model compared to the AUCs obtained by processing the CC and MLO views independently and combining their outputs using mean or maximum operations.

| Model | Two-view model | Mean | Maximum |
|---|---|---|---|
| EfficientNet-B3 (DC) | **0.8643 ± 0.0241** | 0.8420 ± 0.0257 | 0.8426 ± 0.0257 |

#### 3.4.2 | Patch Based Classifier (PBC)

Using the weights obtained from the patch classifier, we constructed single-view patch-based classification (PBC) models and achieved the results presented in Table 14. The best model, ConvNeXt-Base (22k) achieved an AUC of 0.8510 ± 0.0163, which is significantly superior comparing with the second best network, EfficientNet-B3, with a $p$ value of 0.0369, using the hipothesis test from page 6. This training, required approximately 24h for the three rounds on an NVIDIA A100 GPU with 40GB of memory, making it the most time-intensive single-view model in this study.

**TABLE 13** | Classification accuracies of the patch classifiers evaluated on the VinDr-Mammo dataset.

| Model | Test mean ± standard deviation | Best test |
|---|---|---|
| MobileNetV2 | 0.6644 ± 0.0071 | 0.6637 |
| ResNet-50 | 0.6751 ± 0.0130 | 0.6665 |
| DenseNet169 | 0.6793 ± 0.0104 | 0.6798 |
| EfficientNetV2-S (21k) | 0.6857 ± 0.0099 | 0.6970 |
| EfficientNet-B3 | 0.7002 ± 0.0015 | 0.6991 |
| EfficientNet-B0 | 0.6899 ± 0.0092 | 0.7000 |
| ConvNeXt-Base (22k) | 0.6921 ± 0.0133 | **0.7052** |

#### 3.4.3 | Direct Classifier (DC)

The results of the Direct Classifier (DC) experiments are summarized in Table 15. DenseNet169 emerged as the top-performing base model, with an AUC of 0.8134 ± 0.0136, without statistical meaningful superiority ($p$ =0.3985, using the hypothesis test described on page 6), when compared against the second-best network, EfficientNet-B0. Notably, the best-performing Patch-Based Classifier (PBC), ConvNeXt_Base_22k, showed a significant drop in performance when applied to the DC approach. Conversely, DenseNet169, which excelled in the DC task, performed poorly in the PBC setting.

#### 3.4.4 | Conclusions About Single View Classifiers

We compared both results shown above by running the DeLong test and obtained the value $p$ = 0.0013. This indicates with statistical significance that the best PBC model (ConvNeXt-Base (22k)) is superior to the best DC model (DenseNet16) for classifying high-quality mammograms. Due to this result, we recommend the PBC approach for the VinDr-Mammo set with 100% digital mammograms, despite the fact that it requires more training time.

On the other hand, four base models performed better in the Patch-Based Classifier (PBC) approach, while three models showed superior performance in the Direct Classifier (DC) approach (see Figure 5 and Tables 14 and 15). Based on these results, it is not possible to conclude that the PBC approach is universally superior to the DC approach for classifying high-quality mammograms, as performance varies depending on the base model.

**TABLE 14** | AUCs of the single-view classifier based on patches (PBC) for VinDr-Mammo.

| Model | ImageNet input size | Test mean ± standard deviation | Best test | PBC/DC |
|---|---|---|---|---|
| MobileNetV2 | 224 | 0.6978 ± 0.0086 | 0.6858 ± 0.0216 | DC |
| DenseNet169 | 224 | 0.8103 ± 0.0140 | 0.7924 ± 0.0187 | DC |
| EfficientNetV2-S (21k) | 300 | 0.8060 ± 0.0089 | 0.7934 ± 0.0187 | DC |
| ResNet50 | 224 | 0.7797 ± 0.0303 | 0.8005 ± 0.0184 | PBC |
| EfficientNet-B0 | 224 | 0.7986 ± 0.0249 | 0.8208 ± 0.0176 | PBC |
| EfficientNet-B3 | 300 | 0.8172 ± 0.0106 | 0.8280 ± 0.0173 | PBC |
| ConvNeXt-Base (22k) | 224 | 0.8454 ± 0.0086 | **0.8510 ± 0.0163** | PBC |

**TABLE 15** | AUCs of the direct single-view classifier (DC) for VinDr-Mammo.

| Model | ImageNet input size | Test mean ± standard deviation | Best test | PBC/DC |
|---|---|---|---|---|
| ConvNeXt_Base (22k) | 224 | 0.6567 ± 0.0129 | 0.7367 ± 0.0150 | PBC |
| MobileNetV2 | 224 | 0.7734 ± 0.0043 | 0.7653 ± 0.0146 | DC |
| ResNet50 | 224 | 0.7999 ± 0.0042 | 0.7942 ± 0.0140 | PBC |
| EfficientNet-B3 | 300 | 0.8195 ± 0.0092 | 0.8072 ± 0.0137 | PBC |
| EfficientNetV2-S (21k) | 300 | 0.8084 ± 0.0030 | 0.8074 ± 0.0137 | DC |
| EfficientNet-B0 | 224 | 0.8090 ± 0.0067 | 0.8099 ± 0.0136 | PBC |
| DenseNet169 | 224 | 0.8045 ± 0.0067 | **0.8134 ± 0.0136** | DC |

#### 3.4.5 | Conclusions About Base Models

Figure 5 provides a comparison between the performance of the base models in mammogram classification and their performance on ImageNet. The Patch-Based Classifier (PBC) approach exhibits a high correlation coefficient (0.7288), suggesting that its performance closely aligns with the advancements in the evaluated networks. In contrast, the Direct Classifier (DC) approach showed a low negative correlation, indicating a lack of alignment with network improvements. These results suggest that the optimal strategy for classifying high-quality mammograms is to use a modern base model combined with the PBC approach.

### 3.5 | Classifiers With Resized Images in VinDr-Mammo

To answer the Question (3) for VinDr-Mammo, in the same way as the CBIS-DDSM, we downsample mammograms to half their size before classifying them.

Consistent with Section 3.2.3, the best learn-to-resize classifier (LRC; full results in Table 17) underperformed the best fixed-resizing classifier (FRC; full results in Table 16) however without statistical significance, with $p$ value of 0.4388, using the hypothesis test described on page 6. Since LRC is both computationally more complex and resource-intensive than FRC, the Learn-to-Resize approach offers no practical benefit.

In addition, the best classifier with resolution reduction has a substantially lower AUC ($576 \times 448$, $0.8116 \pm 0.0136$) than the best classifier without resolution reduction ($1152 \times 892$, $0.8510 \pm 0.0163$) with statistical significance, with $p$ value of 0.0013, using the hypothesis test described on page 6. Therefore, it is not recommended to reduce the resolution of high-quality digital mammograms before classifying them. This suggests that using an even higher resolution (e.g., $2304 \times 1792$) is likely to improve performance.

### 3.6 | Two-View Classifiers in VinDr-Mammo

To address Question (4), we followed the same network creation and training process outlined in previous sections. We constructed a two-view classifier using the ConvNeXt_Base_22k network, initializing it with weights from the single-view PBC approach. Additionally, we processed each view (CC and MLO) individually using the best-performing classifier from the previous section (ConvNeXt_Base_22k) and combined the results using

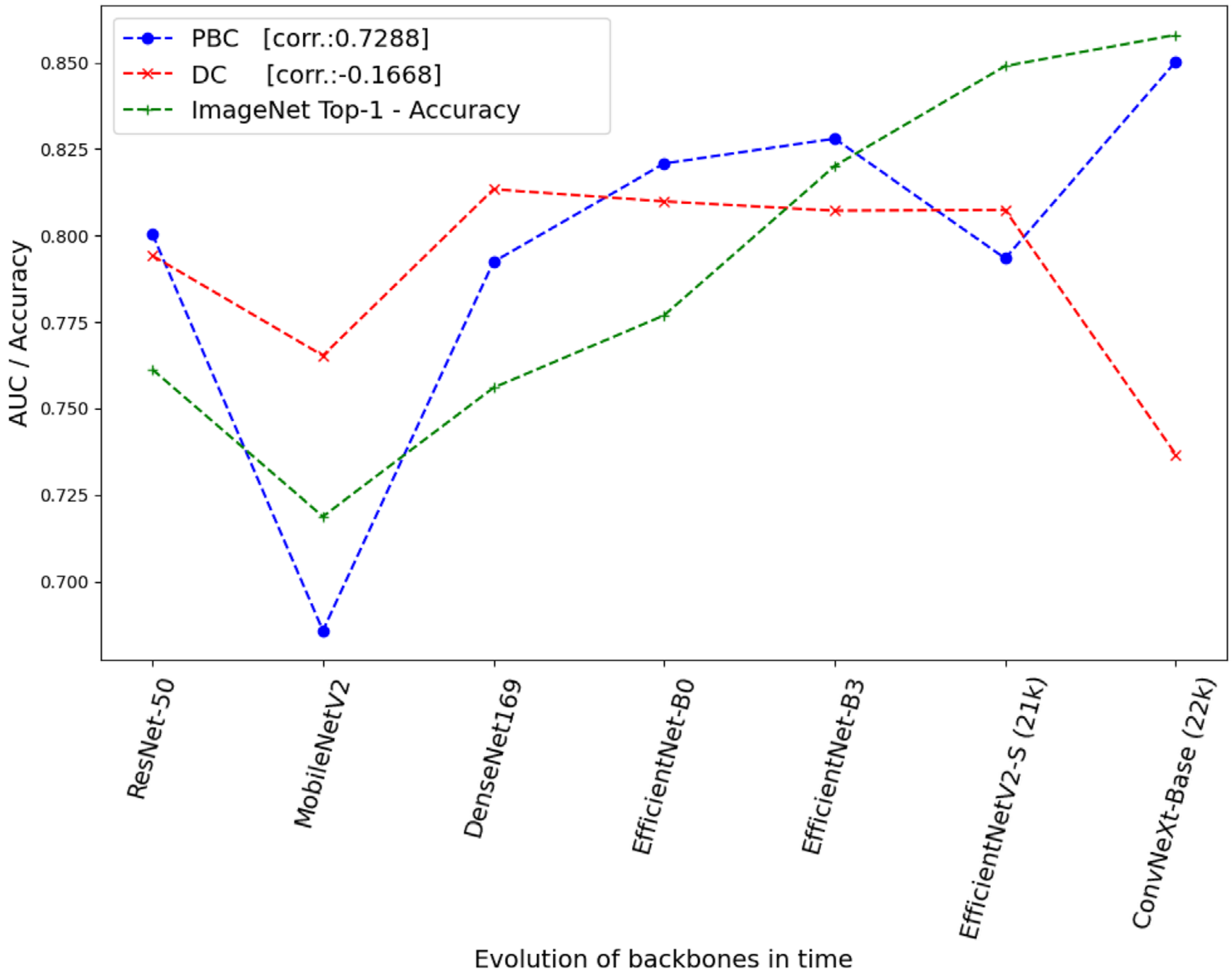

**FIGURE 5** | The AUC scores for all single-view PBC and DC classifiers evaluated on the VinDr-Mammo dataset, accompanied by the ImageNet classification accuracies of their backbone models and the associated Pearson correlation coefficients.

**TABLE 16** | AUCs of fixed resizing classifiers (FRCs) on the VinDr-Mammo dataset.

| Model | ImageNet input size | Test mean ± standard deviation | Best test |
|---|---|---|---|
| MobileNetV2 | 224 | 0.7369 ± 0.0168 | 0.7132 ± 0.0153 |
| EfficientNetV2-S (21k) | 300 | 0.7948 ± 0.0197 | 0.7691 ± 0.0145 |
| EfficientNet-B0 | 224 | 0.7795 ± 0.0025 | 0.7815 ± 0.0143 |
| DenseNet169 | 224 | 0.7903 ± 0.0037 | 0.7856 ± 0.0142 |
| ConvNeXt_Base (22k) | 224 | 0.7642 ± 0.0447 | 0.7869 ± 0.0142 |
| EfficientNet-B3 | 300 | 0.7963 ± 0.0044 | 0.7978 ± 0.0139 |
| ResNet50 | 224 | 0.7916 ± 0.0150 | **0.8116 ± 0.0136** |

**TABLE 17** | AUCs of learn-to-resize classifiers (LRCs) on the VinDr-Mammo dataset.

| Model | ImageNet input size | Test mean ± standard deviation | Best test |
|---|---|---|---|
| ConvNeXt_Base (22k) | 224 | 0.5654 ± 0.0103 | 0.5750 ± 0.0159 |
| DenseNet169 | 224 | 0.6937 ± 0.0945 | 0.7537 ± 0.0148 |
| EfficientNet-B0 | 224 | 0.7640 ± 0.0112 | 0.7622 ± 0.0146 |
| ResNet50 | 224 | 0.7911 ± 0.0172 | 0.7715 ± 0.0145 |
| EfficientNet-B3 | 300 | 0.7869 ± 0.0074 | 0.7767 ± 0.0144 |
| MobileNetV2 | 224 | 0.7589 ± 0.0276 | 0.7910 ± 0.0141 |
| EfficientNetV2-S (21k) | 300 | 0.7891 ± 0.0234 | **0.8095 ± 0.0137** |

**TABLE 18** | AUCs of the two-view model compared to results obtained by averaging and taking the maximum of the CC and MLO views from the single-view model.

| Model | Two-view | Average | Maximum |
|---|---|---|---|
| ConvNext (PBC) | **0.8511 ± 0.0177** | 0.8019 ± 0.0196 | 0.8017 ± 0.0196 |

both average and maximum operations. The results are shown in Table 18.

Applying the per-exam aggregation protocol from Section 2.3.7, we compared the AUC of the two-view classifier with the AUCs obtained by calculating the mean and maximum of the CC and MLO outputs from the single-view classifier. Using the fast version of the DeLong test, we obtained $p$ values of $p = 0.0030$ and $p = 0.0025$, respectively. These results provide strong statistical evidence that the two-view PBC classifier outperforms both the averaging and maximizing approaches applied to the outputs of the single-view classifier.

## 4 | Conclusions

In this study, we systematically evaluated different training approaches for several deep neural network architectures using the film-based CBIS-DDSM dataset (22 architectures from eight network families) and the digital VinDr-Mammo dataset (seven architectures from six families). For low-quality mammograms (CBIS-DDSM), we observed no significant benefit from patch-based pretraining approach; however, it significantly improved classification accuracy on high-quality mammograms (VinDr-Mammo, $p = 0.0013$).

A positive correlation emerged between a model's performance on natural-image classification tasks and mammogram classification when patch-based pretraining was applied (CBIS-DDSM: $r = 0.6491$; VinDr-Mammo: $r = 0.7288$). Conversely, without patch-based pretraining, this correlation was negligible or negative, indicating that modern backbone architectures, like ConvNext for VinDr-Mammo, are advantageous primarily when combined with patch-based pretraining.

Regarding the choice of backbone architecture for mammogram classification, our systematic evaluation yields a dataset-dependent recommendation. For low-quality digitized film mammograms (CBIS-DDSM), EfficientNet-B3 was the top-performing backbone under both patch-based and direct classifier approaches (AUC 0.8325 and 0.8313 respectively), and also produced the best two-view result (AUC 0.8643). Notably, more recent architectures such as ConvNeXt did not surpass EfficientNet-B3 on this dataset, indicating that architectural novelty alone does not translate into better performance when image quality is limited. For high-quality full-field digital mammograms (VinDr-Mammo), ConvNeXt-Base pretrained on ImageNet-22k was the best backbone when combined with patch-based pretraining (AUC 0.8510 single-view, 0.8511 two-view). We also found that ImageNet-22k or -21k pretraining did not consistently outperform ImageNet-1k pretraining, and that base models pretrained on larger ImageNet Input image sizes (300 × 300 or greater) were associated with better results, though the margin was not always statistically significant. Researchers designing new mammographic classifiers can therefore use EfficientNet-B3 as a strong default for film-based or lower-quality datasets, and ConvNeXt-Base with patch-based pretraining as the preferred starting point for FFDM datasets.

Reducing mammogram resolution consistently diminished model performance, especially for high-quality digital images. Furthermore, the "learn-to-resize" technique [35], despite its effectiveness on natural images, did not outperform conventional fixed downscaling on either dataset (CBIS-DDSM: 0.7958 vs. 0.8167; VinDr-Mammo: similar pattern observed). Given its greater computational complexity and lack of performance benefit, learn-to-resize is not recommended for mammogram classification. We hypothesize that the technique's reliance on learned, content-aware resizing is poorly suited to the diffuse, low-contrast nature of mammographic lesions, where uniform downscaling preserves the relevant signal more reliably. Additionally, employing a two-view classification approach substantially outperformed single-view classifiers across both datasets (CBIS-DDSM: $p$ values under 0.0286, while establishing new benchmarks with AUC = 0.8658 ± 0.0239; VinDr-Mammo: $p$ values under 0.0030, achieving an AUC of 0.8511 ± 0.0177). This strongly supports the adoption of multi-view strategies in future mammogram analysis frameworks.

These findings highlight a fundamental three-way trade-off in deep learning–based mammogram classification between image resolution, classification performance, and computational cost. Higher resolutions preserve the fine-grained features that are diagnostically critical, particularly for subtle lesions such as microcalcifications, and consistently yielded the best AUC in our experiments. However, processing high-resolution mammograms requires substantially more GPU memory and training time: the best PBC model on VinDr-Mammo (ConvNeXt-Base at 1152 × 896) required approximately 24 h of training on an NVIDIA A100 (40 GB), and increasing resolution further (e.g., to native 3000 × 4000) is currently impractical even on high-end hardware. Reducing resolution to half (576 × 448) cut memory requirements substantially but significantly degraded AUC on the high-quality VinDr-Mammo dataset (from 0.8510 to 0.8116), confirming that aggressive downscaling is not a viable shortcut for FFDM images. Independently of this resolution trade-off, the patch-based pretraining strategy further improves performance on high-quality digital mammograms by providing domain-adapted initialization weights—learned from lesion-centered crops—that transfer better to whole-image classification than weights pretrained on natural images. Practitioners building mammographic AI systems should therefore plan their resolution and architecture choices jointly with their available hardware, and combine the highest feasible resolution with patch-based pretraining—the most effective configuration our experiments identified.

## Acknowledgments

The Article Processing Charge for the publication of this research was funded by the Coordenação de Aperfeiçoamento de Pessoal de Nível Superior - Brasil (CAPES) (ROR identifier: 00x0ma614).

## Conflicts of Interest

The authors declare no conflicts of interest.

## Data Availability Statement

The data that support the findings of this study are openly available in CBIS-DDSM | Curated Breast Imaging Subset of Digital Database at https://www.cancerimagingarchive.net/collection/cbis-ddsm/ and Vindr-mammo: A large-scale benchmark dataset for computer-aided detection and diagnosis in full-field digital mammography (version 1.0.0) at https://physionet.org/content/vindr-mammo/1.0.0/.